\documentclass[conference]{IEEEtran}
\usepackage{graphicx}
\usepackage[]{algorithm2e}
\usepackage{theorem}
\newtheorem{Theorem}{Theorem}
\newtheorem{lemma}{\textbf{Lemma}}
\vspace{-0.05in}
\begin{document}
\title{Service Composition in Service-Oriented Wireless Sensor Networks with Persistent Queries}
\vspace{-0.1in}
\author{\small
Xiumin Wang$^\dagger$ $^\star$, Jianping Wang$^\star$, Zeyu Zheng$^\star$, Yinlong Xu $^\dagger$, Mei Yang $^\ddagger$\\
$^\dagger$ Department of Computer Science, University of Science and Technology of China, China \\
$^\star$ Department of Computer Science, City University of Hong Kong, Kowloon, Hong Kong\\
$^\ddagger$ Department of Electrical and Computer Engineering, University of Nevada, Las Vegas\\
Email: \{wxiumin2,jianwang\}@cityu.edu.hk,
zzeyu2@student.cityu.edu.hk, ylxu@ustc.edu.cn, meiyang@egr.unlv.edu}

\maketitle \vspace{-0.2in}
\begin{abstract}
Service-oriented wireless sensor network(WSN) has been recently
proposed as an architecture to rapidly develop applications in WSNs.
In WSNs, a query task may require a set of services and may be
carried out repetitively with a given frequency during its lifetime.
A service composition solution shall be provided for each execution
of such a persistent query task. Due to the energy saving strategy,
some sensors may be scheduled to be in sleep mode periodically.
Thus, a service composition solution may not always be valid during
the lifetime of a persistent query. When a query task needs to be
conducted over a new service composition solution, a routing update
procedure is involved which consumes energy. In this paper, we study
service composition design which minimizes the number of service
composition solutions during the lifetime of a persistent query. We
also aim to minimize the total service composition cost when the
minimum number of required service composition solutions is derived.
A greedy algorithm and a dynamic programming algorithm are proposed
to complete these two objectives respectively. The optimality of
both algorithms provides the service composition solutions for a
persistent query with minimum energy consumption.
\end{abstract}
{\bf Keywords:} Service composition, Wireless sensor network,
Routing, Query.

\IEEEpeerreviewmaketitle \vspace{-0.1in}
\section{Introduction}
Service-oriented architecture in WSNs\cite{A.Rezgui,Gracanin} makes
it possible to rapidly develop new applications. In a
service-oriented WSN, a typical application requires several
different services, e.g., data aggregation, data processing,
decoding, which are provided by service providers that are also
sensors. The task of service composition is to assign each required
service to an appropriate service provider according to certain
criteria. Service composition with various performance metrics
\cite{RamanInfocom,Jin,Gu}, e.g., load balance, end-to-end delay and
resource, have been well studied. Service composition in WSNs has
also recently been studied in \cite{Abrams,Srivastava}.
\cite{Abrams} studies the minimum-cost service placement based on
service composition graphs with a tree structure. \cite{Srivastava}
considers the optimal placement of filters (services) with different
selectivity rates.

Habitat and environmental monitoring represent a class of WSN
applications. The queries in such applications in general are {\em
persistent} (or {\em recurrent}) queries which need to be processed
repetitively with a given frequency for a given duration
\cite{A.Rezgui}, e.g., an application requests receiving images in
which the monitored area is dimly lit from 9:00am to
5:00pm\cite{Srivastava}. Three services are required in such a
persistent query: service $s_1$ checking for dim images, service
$s_2$ checking for "sufficient" motion between successive frames,
and service $s_3$ fusing the identified motions(e.g., the appearance
of a suspect). In a service-oriented WSN, such services are provided
by sensor nodes in the network, thus, in-network processing is
feasible, to reduce the possibly massive amounts of raw data.

In WSNs, energy consumption is a critical issue and sleep scheduling
has been well studied as a conservative approach to save energy
\cite{Ha1,wang}. When a node is in sleep mode, all its provided
services are not available, which may cause disruption to service
composition. \cite{wang} studied a cross-layer sleep scheduling
design in a service-oriented WSN which considers system requirement
on the number of active service providers for each service at any
time interval.

\begin{figure}[h]\centering
\includegraphics[width=2.5in]{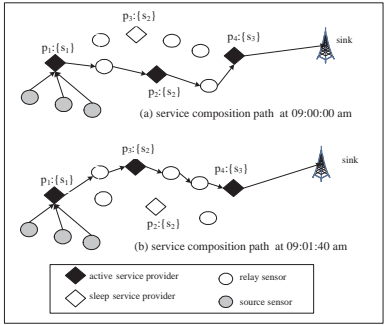}
\caption{ A persistent query requesting $s_1\rightarrow
s_2\rightarrow s_3$ in a service-oriented WSN} \label{example1}
\vspace{-0.065in}
\end{figure}

In a service-oriented WSN, a query routing procedure which routes
requesting services towards service providers is necessary. For a
persistent query, the query routing procedure might need to be
conducted many times during its lifetime due to the sleep schedule
in the MAC layer, which might introduce more energy consumption.
Take the query that starts at 09:00am and ends at 5:00pm with a
frequency of 100s as an example. In Fig.~\ref{example1}(a), at
09:00am, a path is chosen to provide the requested services, while
after 100s one of the sensors in this path switches into sleep mode,
which results in unavailability of the service composition path. It
is necessary to conduct the query routing procedure again to find a
new service composition path as shown in Fig.~\ref{example1}(b). In
this paper, we aim to use the minimum number of service composition
solutions during a persistent query's lifetime such that the energy
consumption caused by repetitively conducting query routing
procedure is minimized. Once the minimum number of required service
composition solutions is derived, we then select the service
composition solutions with minimum transmission cost.

The contribution of the proposed work is summarized as follows:
\begin{itemize}
\item
We propose a service-oriented query routing protocol. Traditional
routing in WSNs only involves finding a path from source sensors to
a sink. Service-oriented query routing protocol needs to ensure that
the path from source sensors to the sink includes service providers,
which imposes new challenges to routing in WSNs.
\item
We propose an optimal greedy algorithm to minimize the number of
required service composition solutions during a persistent query's
lifetime, which can minimize the energy consumption caused by
conducting the service-oriented query routing protocols.
\item
We propose a dynamic programming algorithm to minimize the total
service composition cost which aims to reduce the transmission cost
in executing a query.
\end{itemize}

The rest of the paper is organized as follows. The network
architecture and problem definition are given in Section
~\ref{network} and ~\ref{prob} respectively. The algorithms and
simulation results are presented in Section ~\ref{algorithm} and
~\ref{simulation} respectively. We conclude the paper in Section
~\ref{conclusion}.

\vspace{-0.05in}
\section{Network architecture}
\label{network} 
\begin{figure}[h]\centering
\includegraphics[width=2.7in]{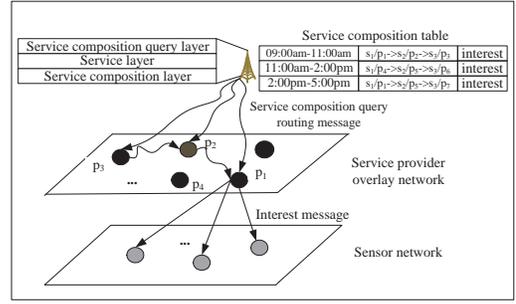}
\caption{System model} \label{model} \vspace{-0.1in}
\end{figure}

In our network architecture, the service providers
form a service provider overlay network as shown in
Fig.~\ref{model}. Two service providers in the service provider
overlay network may be multiple hops away from each other and the
communication between them can be a multi-hop communication in the
same WSN or through existing 802.11 WLAN.

The service-oriented architecture at the sink has
the following three layers:
\begin{itemize}
\item
{\em service composition query layer}. This layer maps an
application's query into a {\em service composition query} which
specifies required services and their invocation order. For example,
the aforementioned query will be converted to a service composition
query with services $s_1$, $s_2$ and $s_3$.
\item
{\em service layer}. This layer has the service information provided
by the sensors in service provider overlay network. We also assume
that service layer has the sleep schedule information of service
providers in service provider overlay network.
\item
{\em service composition layer}. This layer finds the service
composition solutions for service composition queries, which is the
problem to be studied in this paper. For a persistent query, the
service composition layer may find several service composition
solutions during its lifetime since some service composition
solutions may not always be feasible due to sleep schedule. The
service composition solutions are maintained in a {\em service
composition table} as shown in Fig.~\ref{model}.
\end{itemize}
\vspace{-0.03in}

The service composition solution only specifies a service provider
for each required service in a service composition query. Once the
service composition solutions are identified, a routing protocol is
invoked to find paths from source sensors to the first service
provider in the service composition solution and find paths between
any two adjacent service providers. In this paper, we propose the
following service-oriented query routing protocol: \vspace{-0.03in}
\begin{itemize}
\item
The sink broadcasts a {\em service composition query routing}
message which includes service composition solution, duration, and
interest. Such a message will reach all service providers in service
provider overlay network.
\item
Upon receiving a {\em service composition query routing} message, if
a service provider is the first service provider in the service
composition solution, it will broadcast the interest to the sensor
network. Source sensors can then send the data to the first service
provider using any data-driven routing protocol in WSNs. Thus,
service composition is transparent to source sensors.
\item
Upon receiving a {\em service composition query routing} message, if
a service provider is in the service composition solution but not
the first service provider, it needs to find a path to its upstream
service provider in the service composition solution. This can be
done by any routing protocol in WSNs.
\end{itemize}
\vspace{-0.04in}

During the lifetime of a persistent query, it may
be necessary to switch the service composition solutions due to the
sleep schedule of service providers. The service-oriented query
routing protocol needs to be conducted again when the service
composition solution changes, which consumes more energy. The rest
of the paper focuses on the service composition with minimum cost to
avoid the frequent change of service composition solutions during a
persistent query's lifetime.

Notice that the service-oriented query routing protocol is a
distributed routing protocol. The sink only generates the service
composition solutions which determines an appropriate service
provider for each required service. To make such a decision, the
sink only needs to maintain the services availability and the sleep
schedule information of each service provider. In a large-scale WSN,
service providers are only a small portion of the whole network. We
believe that maintaining such information at the sink is well-paid
when the duration of a persistent query is long.
 \vspace{-0.065in}
\section{Problem description}
\label{prob} 
\begin{figure}[h]\centering
\includegraphics[width=2.8in]{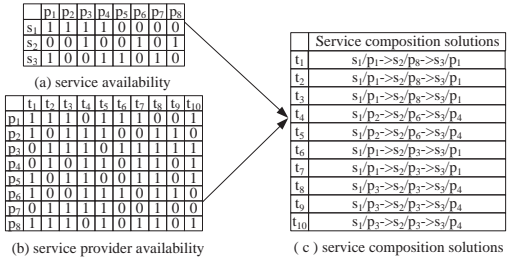}
\caption{An example of service composition for a persistent query
requesting $s_1\rightarrow s_2\rightarrow s_3$ in a service-oriented
WSN} \label{example2}
 \vspace{-0.065in}
\end{figure}

let $S=\{s_1 \rightarrow \cdots \rightarrow s_m\}$ be a
persistent service composition query and $P=\{p_1,\cdots,p_n\}$ be
service providers. Let $S_i$ be the set of services that sensor
$p_i$ can provide and $P_j$ be the set of sensors that can provide
service $s_j$. Fig~\ref{example2}(a) shows the service availability
at the service layer. Given the duration $D$ and the frequency $T$
of a persistent query, the query should be executed for
$\frac{D}{T}$ times during its duration $D$ and we assume that
$\frac{D}{T}$ is an integer. Let $t_k$ be the start time of $k$-th
execution of the persistent query where $1\leq k\leq \frac{D}{T}$.
Given the sleep schedule information of the service providers at the
service layer, the sink can derive each service provider's
availability at $t_k$. Let $x_{ik}$ be 1 if service provider $p_i$
is active at $t_k$, otherwise, set $x_{ik}$ be 0.
Figure~\ref{example2}(b) gives the service provider availability at
the service layer.

With the service availability and the service provider availability
information, the service composition layer can derive a service
composition solution at $t_k$ for $1\leq k \leq \frac{D}{T}$. As
shown in Fig.~\ref{example2}(c), the service composition solution
$s_1/p_1 \rightarrow s_2/p_8 \rightarrow s_3/p_1$ is valid at $t_1,
t_2$ and $t_3$, $s_1/p_2 \rightarrow s_2/p_6 \rightarrow s_3/p_4$ is
valid at $t_4$ and $t_5$ and so on. During this persistent query's
lifetime, 4 service composition solutions are required and thus the
service-oriented query routing protocol needs to be conducted 4
times, which consumes energy. This paper aims to minimize the number
of service composition solutions for a persistent query.

Let $y_k$ be 1 if the service composition solution at $t_k$ is
different from that at $t_{k-1}$, otherwise let $y_k$ be 0. Then
$Y=\sum_{k=1}^{\frac{D}{T}}y_k$ represents total number of service
composition solutions during a persistent query's lifetime, which
needs to be minimized. Under such an objective, a service
composition solution may be used continuously in order to reduce the
energy consumed by frequently invoking service-oriented query
routing protocol. Although some service providers may be used
continuously, this will not decrease the longevity of network. Since
if a service provider is to be active, it has to provide services
for the system according to sleep scheduling.

Though the service-oriented query routing procedure is the major
source of energy consumption, the transmission of the data from the
source sensor to the sink also consumes energy. Two service
providers in the service provider overlay network may be multiple
hops away and if the communication between them is through the same
service-oriented WSN, relay sensors may also be in sleep mode. Thus,
even a service composition solution can be used continuously over
multiple executions, a local routing discovery procedure may be
invoked between two service providers due to the sleep scheduling.
We use average transmission cost between two service providers to
characterize such energy consumption caused by the local routing
discovery between two service providers.
\begin{figure}[h]\centering
\includegraphics[width=2.7in]{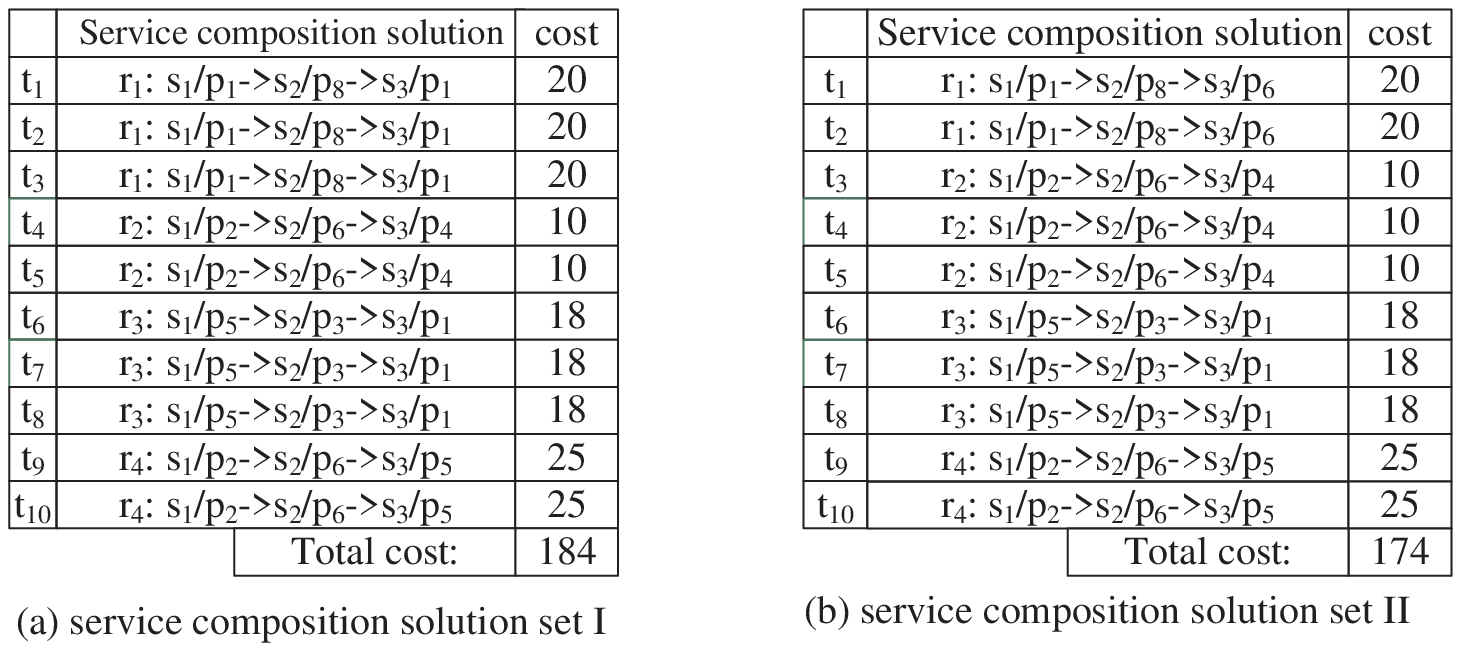}
\caption{Service composition solutions for a persistent query with
the consideration of transmission cost} \label{compositioncost}
 \vspace{-0.065in}
\end{figure}
\vspace{-0.065in}

Besides minimizing the number of service composition solutions
during a persistent query's lifetime, it is also important to
minimize total transmission cost. In Fig.~\ref{compositioncost},
there are two sets of service composition solutions for a persistent
query and Both include 4 service composition solutions during the
persistent query's lifetime. Thus, these two sets of service
composition solutions consumes the same energy caused by
service-oriented query routing procedure. In the first set,
$r_1,r_2,r_3,r_4$ will be used for 3 times, twice, 3 times and twice
respectively with a total cost of 184. In the second set,
$r_1,r_2,r_3,r_4$ will be used for twice, 3 times, 3 times and twice
respectively with a total cost of 174. Thus, the second set of
service composition solutions will be more energy efficient.

In this paper, firstly, we aim to minimize the number of service
composition solutions during a persistent query's lifetime. Such a
problem is referred to as problem {\bf P1}. Secondly, we need to
minimize the total cost of the service composition solutions. Such a
problem is referred to as problem {\bf P2}.
 \vspace{-0.07in}
\section{Algorithm design and analysis}\vspace{-0.02in}
\label{algorithm} In this section, we first approach problem {\bf
P1}. Then based on the result of {\bf P1}, we approach the second
problem {\bf P2}.
 \vspace{-0.1in}
\subsection{Greedy algorithm for problem {\bf P1}}
 \vspace{-0.05in}
Let $avl_{ik}$ be the number of executions that service provider
$p_i$ can be continuously available from $k$-th execution (including
at $k$-th execution). For example, if $p_i$'s availability at all
execution instances of a persistent query is given as $1100111001$,
$avl_{i1}$ is 2 since $p_i$ can be available at $1$st and $2$nd
execution, $avl_{i3}$ is 0 as $p_i$ is not available at $3$rd
execution.

The greedy algorithm which is shown in Algorithm.~\ref{greedy} is
always to select the service provider with maximum $avl_{ik}$ for
each $s_j$ in $k$-th execution such that the solution can be
continuously used for the maximum number of times. After the service
composition solution is determined for $k$-th execution, the number
of times that this solution can be used is determined by the minimum
$avl_{ik}$ among all selected service providers. Let $SC_k$ be the
set of selected service providers for $k$-th execution and $num_h$
be the number of times that $h$-th service composition solution can
be continuously used.

The worst case running time of this greedy algorithm is
O($\frac{D}{T}\*m\*n$). $\sum_{k=1}^{\frac{D}{T}}y_k$ gives the
minimum number of service composition solutions during a persistent
query's lifetime.
We now prove the optimality of the greedy algorithm. Let
$Y=\sum_{k=1}^{\frac{D}{T}}y_k$ be the solution obtained from the
greedy algorithm where $y_{l_1}=1, y_{l_2}=1,\cdots, y_{l_Y}=1$. Let
$Y'$ be an optimal solution where
$y_{l'_1}=1,y_{l'_2}=1,\cdots,y_{l'_{Y'}}=1$.
\vspace{-0.1in}
\begin{lemma} \label{lemma}
For any sequence $l_1,\cdots, l_b, \cdots, l_r$ and $l'_1,\cdots,
l'_b, \cdots, l'_{r}$ where $1\leq r \leq min\{Y,Y'\}$, there must
always exists $l_1\geq l'_1,\cdots, l_b\geq l'_b, \cdots, l_r\geq
l'_r$.
\end{lemma}
\vspace{-0.08in}
\begin{proof}
We use induction to prove this lemma. Firstly, for $b=1$, it is
obvious that $l_1=l'_1=1$. When $b=2$, as greedy algorithm always
selects the provider with maximum $avl_{ik}$ for each service, the
value of $(l_2-l_1) -(l'_2-l'_1)$ must be no less than 0, so
$l_2\geq l'_2$. Assume that when $b=d$ we have $l_d\geq l'_d$. For
$b=d+1$, in the given optimal solution, there is a service
composition solution which can be continuously used from $l'_d$ to
$l'_{d+1}$. If $l'_{d+1} \leq l_d$, then we have $l_{d+1} \geq l_d
\geq l'_{d+1}$; If $l'_{d+1} > l_d$, then we must have $l_{d+1} \geq
l'_{d+1}$ since the greedy algorithm always selects the service
providers which can provide longest continuous services. In both
cases, we have $l_{d+1} \geq l'_{d+1}$. Thus, lemma holds when
$b=d+1$.
\end{proof}
\vspace{-0.1in}
 \restylealgo{algoruled}
\begin{algorithm}[]
 \linesnumbered
 \scriptsize
{
 \Begin{
 $avl_{ik}=0$, $SC_k=\emptyset$ and $y_k=0$ where $1\leq h\leq \frac{D}{T}$, $1\leq i\leq n$, $1\leq k\leq \frac{D}{T}$\;$h=k=1$\;
 \For{$k=1$ to $\frac{D}{T}$} { \For{$p_i \in P$} {
     calculate the value of $avl_{ik}$\;
} }
 \While{$k \leq \frac{D}{T}$}{
    \For{each service $s_j \in S$}{
      $SC_k = SC_k \cup \arg \max_{p_i \in P_j}\{avl_{ik}\}$.\
    }
  $num_h=\min_{p_i\in SC_k}\{avl_{ik}\}$\;
  $y_k$=1;
  $k=num_h+k$;
  $h=h+1$\;
  }
  }
\caption{Greedy algorithm}\label{greedy}}
\end{algorithm}
\vspace{-0.1in}
\begin{Theorem}\label{theorem1}$Y$, the solution obtained from the greedy algorithm, must be optimal.
\vspace{-0.04in}\end{Theorem} \vspace{-0.04in}
\begin{proof}
We prove it by contradiction. Assume that there exists $Y>Y'$, then
$l_{Y}> l_{Y'}$. According to lemma 1, we also have $l_{Y'}\geq
l'_{Y'}$. The relationship among $l_Y$, $l'_{Y'}$, and $l_{Y'}$ is
shown in Fig.~\ref{compare}.
\begin{figure}[h]\centering
\includegraphics[height=0.6in]{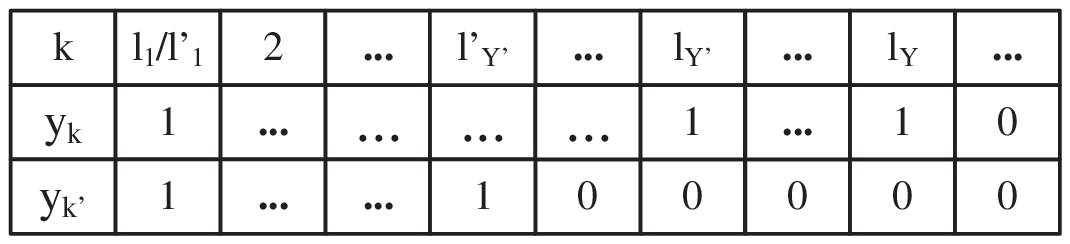}\vspace{-0.05in}
\caption{The value of $y_k$ and $y_{k'}$} \vspace{-0.05in}\label{compare}
\vspace{-0.1in}
\end{figure}
Fig.~\ref{compare} denotes that there exists a service composition
solution which can cover executions from $l_{Y'}$ to $l_Y$.
According to our greedy algorithm, we can find a solution which can
be continuously used from $l_{Y'}$-th execution to the last
execution. Thus $Y=Y'$, which conflicts the assumption.
\end{proof}
 \vspace{-0.12in}
\subsection{Minimize the total service composition cost }
In the following, we approach problem {\bf P2} which minimizes the
total routing cost based on the result of problem {\bf P1}.

Let $Sol_{k,q}$ be the set of feasible service composition solutions
at $k$-th execution which can be continuously used for the following
$q$ executions. For any service composition solution $X\in
Sol_{k,q}$, let $c_{k,q}(X)$ be transmission cost if $X$ is selected
to be executed once. Let $C(k,q)=\min_{X\in Sol_{k,q}} c_{k,q}(X)$.
$C(k,q)$ can be obtained by finding a shortest path in an auxiliary
graph $G=(V; E)$ which is constructed as follows:
\begin{itemize}
\item
$V$ is the set of nodes consisting of $m$ layers $V_1,\ldots, V_m$
and the $j^{th}$ layer $V_j$ contains all service providers which
can continuously provide  $s_j$ from $k$-th execution to $(k+q)$-th
execution, e.g.,  if $p_i$ can provide $s_j$ and it is available
from $k$-th execution to $(k+q)$-th execution, node $v_{ji} \in
V_j$.
\item
Let $E$ be the link set such that there is a direct link $e_{j-1,i,
j,h}\in E$ whenever $v_{(j-1)i}\in V_{j-1}$ and $v_{jh}\in V_{j}$
for $j\in \{2,\ldots, m\}$. The cost of $e_{j-1,i,j,h}$ is the
shortest path cost from $p_i$ to $p_h$ in the physical network.
\item
Add two special nodes $s$ and $d$ such that $\{s\}$ is the $0^{th}$
layer and $\{d\}$ is the $(m+1)^{th}$ layer. Link $s$ to each node
in $V_1$ and link each node in $V_m$ to $d$ with cost 0.
\end{itemize}
\vspace{-0.05in}
\begin{algorithm}[]
 \linesnumbered
\scriptsize {
 \Begin{
 \For{$k=Y$  to $\frac{D}{T}$}
  {
  $cost(k,Y)=C(k, \frac{D}{T}-k+1)*(\frac{D}{T}-k+1)$\;
  }
  \For{$h=Y-1$ downto 1}
  {
   \For{$k=h$ to $(\frac{D}{T}-Y+h)$}
   {
     $cost(k,h)=min_{q\in [1,\frac{D}{T}]}(C(k,q)*q+cost(k+q,h+1))$\;
     $sw[h,k]=q$\;
   }
  }
  $k=1$\;
  \For{$h=1$ to Y}
  {
    $times=sw[h,k]$\;
    $route[h] \leftarrow$ the service composition solution with minimum $C(k, times)$ \;
    $k=k+times$\;
    }
 }
  \caption{Dynamic programming algorithm} \label{dynamic}
  }
  \vspace{-0.04in}
\end{algorithm}
\vspace{-0.09in}

Let $cost(k,h)$ be the minimum total cost from $k$-th execution to
the last execution if $h$-th service composition solution starts at
$k$-th execution. Then we have the following recursion:
{\small $$cost(k,h)=min_q(C(k,q)*q+cost(k+q,h+1))$$}
\vspace{-0.05in}
where $1\leq h\leq Y,h\leq k\leq \frac{D}{T}-Y+h$.
We have the following boundary condition:
{\small $$cost(k,Y)=C(k,\frac{D}{T}-k+1)*(\frac{D}{T}-k+1)$$}
\vspace{-0.02in} for $k=Y, \ldots,\frac{D}{T}$.

The dynamic programming is given in Algorithm.~\ref{dynamic} in
which $cost(1,1)$ is the minimum total cost for the persistent query
and $route[h]$ stores $h$-th service composition solution. The time
complexity of the algorithm is $O((\frac{D}{T})^3)$.

\vspace{-0.1in}
\section{Simulation results}
In this section, we first introduce the design of our simulation.
The number of service providers $d_j$ of each service is randomly
generated between $[15\%n,25\%n]$. We then randomly generate $d_j$
service providers for $s_j$ from $p_1, p_2,\ldots, p_n$. For each
$p_i$, we also randomly generate its availability at each execution.
Then we validate whether each $s_j$ can be provided by at least one
active service provider at each execution. If infeasible, the
instance is dropped from our simulation.

To compare the performance of our algorithms, we also introduce a
baseline algorithm named {\em min-cost-based} algorithm which aims
to select the service composition solution with minimum transmission
cost for each execution. We compare the number of service
composition solutions during a persistent query's lifetime and the
total transmission cost of our algorithms with {\em min-cost-based}
algorithm respectively. \label{simulation}
\begin{figure}[h]\centering
\includegraphics[height=1.6in]{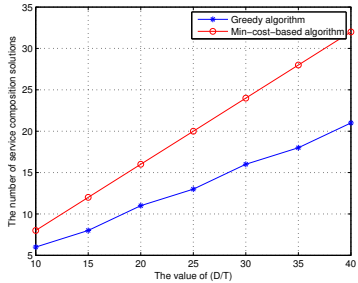} \caption{Minimum number of service composition solutions
 with greedy algorithm versus the value of $\frac{D}{T}$} \label{solutions}
\vspace{-0.1in}
\end{figure}

In the first set of experiments, we evaluate the performance of
algorithms by varying $\frac{D}{T}$ in $\{10,15,20,25,30,35,40\}$
for $m=20,n=40$. The effectiveness of greedy algorithm for {\bf P1}
is tested by comparing with {\em min-cost-based} algorithm. As shown
in Fig.~\ref{solutions}, the number of service composition solutions
during persistent query's lifetime in greedy algorithm is much less
than that in {\em min-cost-based} algorithm. For example, with
$\frac{D}{T}=40$, the number of service composition solutions in our
greedy algorithm is only 21 while it is 32 in {\em min-cost-based}
algorithm. The difference between two algorithms increases with the
number of query's executions, which demonstrates the effectiveness
and scalability of our work.

\begin{figure}[h]\centering
\includegraphics[height=1.6in]{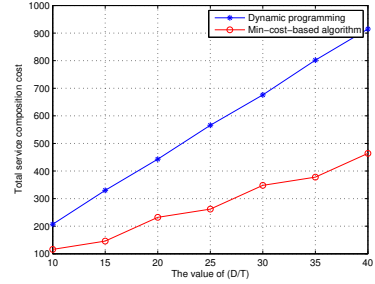} \caption{Total service composition cost versus the value of $\frac{D}{T}$} \label{cost1}
\vspace{-0.1in}
\end{figure}

Fig.~\ref{cost1} illustrates that total service composition cost
obtained from dynamic programming based on the result of the greedy
algorithm is higher than that obtained from {\em min-cost-based}
algorithm. As we explained in section~\ref{prob}, the energy
consumed in service-oriented query routing protocol is much higher
than that in conducting service composition. Thus, though the
solution obtained from our algorithms may consume more energy in the
service composition phase, it consumes much less energy in
service-oriented query routing phase which is the major energy
consumption source in a persistent query.

\begin{figure}[h]\centering
\includegraphics[height=1.6in]{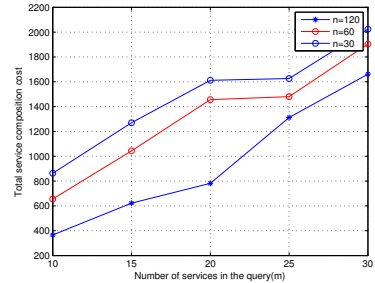} \caption{Total cost in each network topology with dynamic programming versus the number of services needed to provide} \label{cost2}
\vspace{-0.10in}
\end{figure}

In the second set of our experiments, we study in detail the impact
of the number of required services on the total service composition
cost and the impact of the number of service providers on the total
service composition cost. We have selected three scenarios
$(n=120,\frac{D}{T}=40)$, $(n=60,\frac{D}{T}=40)$,
$(n=30,\frac{D}{T}=40)$ by varying $m$ in $[10, 30]$. As shown in
Fig.~\ref{cost2}, the total service composition cost increases with
the number of  required services $m$ since more service providers
may be involved in a service composition. Given $m$, the service
composition cost is lower in a network with more service providers.
In a network with more service providers, more feasible service
composition solutions are possible and our dynamic programming
algorithm can find the service composition solution with minimum
cost.
\vspace{-0.05in}
\vspace{-0.15in}
 \vspace{-0.135in}
\section{Conclusion}
\vspace{-0.08in} \label{conclusion} This paper studies service
composition in service-oriented WSNs with persistent queries. We aim
to provide service composition solutions during a persistent query's
lifetime such that the involved routing update cost and transmission
cost is minimized. The optimality of greedy algorithm and dynamic
programming provides the service composition solutions for
persistent queries with the minimum energy consumption.

\end{document}